\documentclass[prd,nofootinbib,floats,superscriptaddress,eqsecnum,tightenlines,11pt]{revtex4}

\usepackage{hyperref}
\usepackage{graphicx}
\usepackage{amsmath,amssymb,amsfonts,amsthm,latexsym,stmaryrd}
\usepackage{marginnote}
\usepackage{color}

\def\beq{\begin{equation}}
\def\eeq{\end{equation}}
\newcommand{\bea}{\begin{eqnarray}}
\newcommand{\eea}{\end{eqnarray}}
\def\bi{\begin{itemize}}
\def\ei{\end{itemize}}
\def\ba{\begin{array}}
\def\ea{\end{array}}
\def\bfig{\begin{figure}}
\def\efig{\end{figure}}

\begin{document}

\title{Semi-classical analysis of black holes in Loop Quantum Gravity: \\
Modelling Hawking radiation with volume fluctuations}
\author{P. Heidmann}
\email{pierre.heidmann@ens-lyon.fr}
\affiliation{CEA, IPhT, 91191 Gif-sur-Yvette, France}
\affiliation{D\'epartement de Physique, \'Ecole Normale Sup\'erieure de Lyon, Universit\'e de Lyon, 46 All\'ee d'Italie, 69007 Lyon, France}
\author{H. Liu}
\email{hongguang.LIU@etu.univ-amu.fr}
\affiliation{Centre de Physique Th\'eorique (UMR 7332) , Aix Marseille Universit\'e and Universit\'e de Toulon, 13288 Marseille, France}
\affiliation{Laboratoire de Math\'ematiques et Physique Th\'eorique, Universit\'e Fran\c cois Rabelais, Parc de Grandmont, 37200 Tours, France}
\affiliation{Laboratoire APC -- Astroparticule et Cosmologie, Universit\'e Paris Diderot Paris 7, 75013 Paris, France}
\author{K. Noui}
\email{karim.noui@lmpt.univ-tours.fr}
\affiliation{Laboratoire de Math\'ematiques et Physique Th\'eorique, Universit\'e Fran\c cois Rabelais, Parc de Grandmont, 37200 Tours, France}
\affiliation{Laboratoire APC -- Astroparticule et Cosmologie, Universit\'e Paris Diderot Paris 7, 75013 Paris, France}

\date{\today}

\begin{abstract}
We introduce the notion of fluid approximation of a quantum spherical black hole in the context of Loop Quantum Gravity. In this 
limit, the microstates of the black hole are intertwiners between ``large'' representations $s_i$ which typically 
scale as $s_i \sim \sqrt{a_H}$  where $a_H$ denotes the area of the horizon in Planck units.
The  punctures with large colors 
are, for the black hole horizon, similar to what are the fluid parcels for a classical fluid. We dub them puncels.
Hence, in the fluid limit, the horizon is composed by puncels which are themselves interpreted as
composed (in the sense of the tensor product) by a large number of more fundamental intertwiners.
We study the spectrum of the euclidean volume acting on  puncels and we compute its quantum fluctuations. 
Then, we propose  an interpretation of  black holes radiation based on the properties of the 
quantum fluctuations of the euclidean volume
operator. We estimate a typical temperature of the black hole and we show that it scales as 
the Hawking temperature. 
\end{abstract}

\maketitle

\section{Introduction}
The last twenty years,  several approaches have been developed 
in the framework of Loop Quantum Gravity to describe black hole microstates 
which are responsible for their huge entropy and to explain more generally 
their famous thermodynamical properties \cite{Bekenstein,Hawking}.
After the first heuristic but seminal model of black holes in Loop Quantum Gravity \cite{Rovelli}, 
one introduced the idea of isolated horizons \cite{Ash0, Ash1, Ash2, Ash3} which allowed for a very clear 
description of  the black hole microstates in terms of a Chern-Simons theory \cite{Ash4,Ash5,BHentropy8, BHentropy6}.
Most of these models have in common that counting the microstates leads to recovering the Bekenstein-Hawking law only when the Barbero-Immirzi parameter $\gamma$ is fixed to a special value. 
The fact that $\gamma$ plays such an important role has raised some doubts and criticisms about the way
we describe black holes in Loop Quantum Gravity. However, a new perspective has been opened  the last
years \cite{BHentropy8, BHentropy6} showing that it is possible to fix  $\gamma = \pm i$  to compute black holes
entropy which reproduces exactly the Bekenstein-Hawking law at the semi-classical limit. 
This result suggests that the quantum theory, when described in
terms of self dual variables \cite{Ashtekar:1986yd}, might automatically account
for a right description of quantum black holes. See \cite{Alereview} for a complete recent review on black holes in Loop
Quantum Gravity.

If the entropy and the microstates of black holes has been studied a lot, very few models about  Hawking radiation in Loop Quantum Gravity 
were developed. The quasi-local approach \cite{GhoshAle1,GhoshAle2} allowed to introduce the notion of temperature; 
in \cite{Bianchi rad}, one made used of this temperature to develop a model of radiation in the context of Spin-Foam models; 
in \cite{radiation} radiation has been explained from a group theoretical point of view in canonical Loop Gravity. 
However, none of these models really propose  a full and dynamical fundamental explanation of the Hawking radiation. 
We propose and explore an idea to make progress in this direction.
It consists in assuming that the black hole radiation originates
from quantum fluctuations of its horizon. Hence, we study some properties of these fluctuations at the semi-classical level from the properties of the
spectrum of the euclidean volume operator
acting on the space of black hole microstates. Then, 
we show that they could be interpreted in terms of defects spread on the horizon, and the typical length $\lambda$ between 
two such defects  scales as $\sqrt{a_H}$ where $a_H$ is the area of the horizon. Then, we argue that  $\lambda$ could be viewed as a fundamental
wavelength of the black hole and one shows that
it corresponds to a temperature $T$ which naturally scales as the Hawking temperature. 

To construct this model, first we introduce the concept of fluid approximation for a black hole in Loop Quantum Gravity. This approximation consists in
considering only a particular class of ``semi-classical" states for the black hole. This class is such that the representations coloring the punctures which cross the horizon are 
semi-classical (they scale as $\sqrt{a_H}$ in Planck units according to \cite{Noui2}), 
and the intertwiners describing the microstates are restricted  to a particular cathegory. 
These intertwiners   are associated to a graph $\Gamma$ 
dual to a triangulation $\Delta$ of the ball as shown in figure \ref{puncels}. Hence, they are obtained by contracting together $n$ 4-valent intertwiners between large 
representations. These 4-valent intertwiners are for the black hole what parcels are for a fluid, and we dub them puncels.  
To understand thermodynamical aspects of the black hole in the fluid limit, we proceed as one usually does for the calculation of black holes path integral:
we transform the lorentzian geometry into a euclidean one, hence the black hole is transformed into a 3-ball whose boundary is the horizon. 
We show that the properties of the quantum geometry of this 3-ball allows us to interpret the Hawking radiation from the point of view of Loop Quantum Gravity. 
Indeed, we compute the semi-classical
spectrum of the euclidean  volume operator acting on the semi-classical states, and we interpret volume fluctuations in terms of deformations of the 
black hole horizon. Finally, as we said above, we can easily interpret these deformations as defects which are  spread on the horizon 
with a typical length $\lambda \sim \sqrt{a_H}$.
Note that we mostly work in the model where $\gamma = \pm i$ (the representations are continuous) but our analysis could be adapted 
to the case $\gamma \in \mathbb R$.

The paper is organized as follows. Section II is devoted to define the fluid approximation of the black hole. In Section III, we compute the spectrum of the volume
operator in the fluid approximation, and we relate the volume fluctuations to the Hawking radiation. We conclude in Section IV with a discussion on this model.

\section{Black Holes in the fluid approximation}
In this section, we introduce the concept of fluid limit of a spherical black hole. We first recall some
fundamental properties of  $SU(2)$ black holes in the context of Loop Quantum Gravity. Then,
we define their  fluid approximation. 

\subsection{Black Holes and $SU(2)$ Chern-Simons theory}
It is well-known that the quantum microstates of a spherical black hole are those of an $SU(2)$ Chern-Simons theory
on a punctured 2-sphere $S^2$ \cite{BHentropy8,BHentropy6,BHentropy7}. Let us quickly recall how this works. 

\subsubsection{Microstates are intertwiners}
Even though $S^2$ is a topological sphere (with no geometrical structure), it  represents the horizon of the black hole
from a classical point of view. The punctures have a quantum nature and originate from those edges of quantum geometry 
states which cross the horizon. They are a priori colored with spins $j$ and they carry quanta of area $a_j=8\pi \gamma \ell_p^2 \sqrt{j(j+1)}$
where $\gamma$ is the Barbero-Immirzi parameter and
$\ell_p$ the Planck length.  Hence, the Hilbert space
of quantum black hole microstates is given by
\bea\label{space of states}
\bigoplus_{(n_j) \in \, {\cal C}(a_H)} \text{Int}(\bigotimes_{j=1}^\infty V_{j}^{\otimes n_j})
\eea
where $V_j$ denotes the (modulus of the) spin $j$ representation of dimension $d_j=2j +1$, $n_j$ is the number of punctures colored with $j$,
and we have introduced the constraint ${\cal C}(a_H)$ 
\bea\label{area}
a_H - 8\pi \ell_p^2 \gamma \sum_{j} n_j \sqrt{j(j+1)}= 0
\eea
which ensures that the area of the horizon is fixed to $a_H$.   In fact, this description of the space of microstates has already taken into
account that the black hole is semi-classical in the sense that $a_H \gg \ell_p^2$. When the black hole is ``smaller'', we must consider
quantum corrections. In fact, classical $SU(2)$ intertwiners 
are replaced by quantum intertwiners, more precisely by intertwiners of the quantum group $U_q(su(2))$ where $q$ is a root of unity. Details can be found in \cite{BHentropy7}. 

\subsubsection{Number of microstates and semi-classical limit: $\gamma \in \mathbb R$}
The dimension ${\cal N}(a_H)$ of the space of states \eqref{space of states} can be computed easily from group theory techniques. It is formally given by
\bea\label{micro}
{\cal N}(a_H) = \sum_{ {\cal C}(a_H)} \text{dim}[\text{Int}(\bigotimes_{j=1}^\infty V_{j}^{\otimes n_j})]
\eea
where the sum runs over the configurations of spins $(n_j)$ which satisfy ${\cal C}(a_H)$ and
\bea
\text{dim}[\text{Int}(\bigotimes_{j=1}^\infty V_{j}^{\otimes n_j})]  \; = \; \frac{2}{\pi} \int_0^\pi d\theta \, \sin^2 \theta \, \prod_{j} \left( \frac{\sin(d_j \theta)}{\sin \theta}\right)^{n_j}
\eea
is the number of invariant tensors in the tensor product $\otimes_{j} (V_{j}^{\otimes n_j})$. Many different methods have been
developed  to compute the first terms of the 
semi-classical expansion ($a_H \gg \ell_p^2$)  of  ${\cal N}(a_H)$ \cite{Rovelli,kaulma1,kaulma2,kaulma3,counting1,counting2,counting3,counting4,counting5,BHentropy3,Liv1,Bianchi}. One shows that it
reproduces the Bekenstein-Hawking law at the leading order provided that $\gamma$ is fixed to a particular numerical value $\gamma_0$, i.e.
\bea\label{semi classical}
S(a_H) \equiv \ln {\cal N}(a_H) = \frac{\gamma_0}{\gamma} \frac{a_H}{4\ell_p^2} - \frac{3}{2} \ln \left( \frac{a_H}{\ell_p^2}\right) + \cdots
\eea
The explicit  value of $\gamma_0$ can be computed numerically (see \cite{BHentropy7} for example).
Note that the punctures are assumed to behave as distinguishable particles to obtain the Bekenstein-Hawking law.
Furthermore, the semi-classical limit can be shown to be dominated by configurations where the spins are small in the sense
that the mean color and then the mean number of punctures when the black hole is macroscopic are given by:
\bea
{1}/{2} < \langle j \rangle < 1  \qquad \text{and} \qquad \langle n \rangle = n_0 \frac{a_H}{\ell_p^2} \, ,
\eea
where $n_0$ is a constant. The means are computed in the microcanonical ensemble according to the partition function
\eqref{micro}.

\subsubsection{Analytic continuation to $\gamma=\pm i$}
It goes without saying that the strong dependence of the black hole entropy computation  on $\gamma$ has
remained  a  controversial  aspect. The first reason is that the Barbero-Immirzi parameter plays no role at the semi-classical level
whereas it is crucial at the quantum level, at least in the understanding of black holes entropy. Second,  the value
of $\gamma_0$ \eqref{semi classical} for some models of rotating black holes is different from the one computed in the case of spherical black holes
\cite{Ale2,rotating}. All these results strongly suggest to look at alternative methods to understand thermodynamical properties of
black holes in Loop Quantum Gravity. 

A first promising perspective on the issue of the dependence of black holes entropy in $\gamma$ was put forward
thanks to the availability of the canonical ensemble formulation  of  the  entropy  calculation  making  use  of  the
quasi-local description of black holes \cite{GhoshAle1,GhoshAle2,GhoshAle3}. 
Before, most of the calculations had been performed in the microcanonical ensemble and then, they were
reduced to a calculation of number of microstates. The introduction of a notion of quasi-local energy $E(n_j)$ for black holes microstates enabled us to
compute a canonical (and eventually a grand canonical) partition function ${\cal Z}(\beta)$ where $\beta$ is the inverse temperature \cite{GhoshAle3}. 
It is 
formally defined by
\bea
{\cal Z}(\beta) = \sum_{(n_j)} \mu(n_j) \exp[- \beta E(n_j)]
\eea
where the sum runs over all possible microscopic configurations $(n_j)$ satisfying ${\cal C}(a_H)$. The  measure $\mu$ takes into account the
matter sector and the symmetrisation factors.
 It was shown in \cite{GhoshAle3} that the semi-classical analysis of the partition function allows to recover the right semi-classical behavior for all values of
$\gamma$.  The semi-classical limit occurs at the vicinity of the Unruh temperature $\beta_u$ (defined for the ``quasi-local" observer
in the context of the ``quasi-local" description of
black holes) where the entropy behaves as
\bea
S(\beta_u) =  \ln {\cal Z}(\beta_u) - \beta_u \frac{\partial \ln {\cal Z}}{\partial \beta}(\beta_u) =  \frac{a_H}{4\ell_p^2} + \cdots
\eea
More precisely, the entropy of large semiclassical black holes coincides with the Bekenstein-Hawking law  at the leading order, while 
the dependence on the Immirzi-Barbero parameter is shifted to sub-leading quantum corrections. In this approach, the semi-classical limit is dominated
by large spins in the sense that the mean color and the mean number of punctures generically scale as 
\bea\label{means}
\langle j \rangle \approx \sigma \frac{\sqrt{a_H}}{\ell_p} \, \qquad \text{and} \qquad
\langle n \rangle \approx \nu \frac{\sqrt{a_H}}{\ell_p} \, .
\eea
Here $\sigma$ and $\nu$ are constant, and the means are computed in the grand canonical ensemble. 
We assumed that the punctures satisfy the Bose-Einstein or the Maxwell-Boltzmann statistic in the semi-classical regime. A similar result
exists for the Fermi-Dirac statistic \cite{GhoshAle3}.

Another perspective to understand the semi-classical limit of spherical black holes which relies on analytic continuation techniques
was  originally proposed in \cite{Noui1}. This idea was  developed further in \cite{Noui2,Noui0,JBA1}. It was 
adapted to three-dimensional BTZ black holes in \cite{BTZ},  
it was recently generalized to models of rotating black holes in \cite{rotating}, and it was also applied to explain Hawking radiation \cite{radiation}
 from group theory
arguments. 
Note that this idea is strongly supported by different complementary approaches in different contexts \cite{J1,J2,Pranz1,Pranz2,Pranz3,Ale1,Yasha1,Yasha2,BN,Muxin}. 
In all these cases, the number of microstates (viewed as a function of the parameter $\gamma$) is analytically continued  
from $\gamma \in \mathbb R$ to  $\gamma \in \mathbb C$  and then evaluated at the special complex values $\gamma=\pm i$.
A simple analysis shows that it grows asymptotically as $\exp(a_H/(4\ell_p^2))$ for large areas. 
The result is striking in that these values of the Immirzi parameter are special in the connection formulations of gravity:
they lead to the simplest covariant parametrization of the phase space of general relativity in terms of the so-called Ashtekar variables \cite{Ashtekar:1986yd}. 
Moreover, this suggests that the quantum theory, when defined in terms of self dual variables,  might automatically account for a holographic degeneracy of the area spectrum of the 
black hole horizon. The main differences with the real $\gamma$ description of black holes is that, first, the horizon area has now a continuous spectrum, and second the 
semi-classical limit is  dominated by large representations. 
Indeed, each spin $j$ coloring  the punctures are mapped into the complex number $j \mapsto 1/2(-1 + is)$ which depends on the real parameter $s$. 
In that way,
quanta of area remain real when $\gamma$ takes the values $\gamma= \pm i$
\bea\label{analytic rules}
a_j=8\pi \gamma \ell_p^2 \sqrt{j(j+1)} \longmapsto 
a_j = 4\pi \ell_p^2 \sqrt{s^2 + 1} \qquad \text{with} \quad \gamma = \pm i \, , \,\, j=  \frac{1}{2}(-1 + is) \, .
\eea
Generically (when the particles satisfy the Maxwell-Boltzmann statistics at the semi-classical limit), the mean representation and the mean number of punctures  were shown to scale
exactly as \eqref{means}  in the semi-classical regime \cite{Noui2}. Hence, these conclusions coincide with the analysis of the partition function using the quasi-local energy \cite{GhoshAle3} (briefly recalled  above) with the difference that discrete spins are replaced by continuous representations.  It is interesting
to emphasize once more that the analytic continuation method was successfully adapted to models of rotating black holes \cite{rotating}: the entropy
of a rotating black hole reproduces the expected semi-classical law when $\gamma = \pm i$ whereas this is clearly not the case when $\gamma$ is real.

\subsection{Fluid approximation and fluid ``puncels"}
From now on, we consider the model based  on the analytic continuation where the semi-classical limit is dominated by large continuous
representations.  This model is more satisfying than the others mainly because it
leads to a $\gamma$-independent semi-classical behavior of the black hole entropy (at least at the leading order). Furthermore, contrary to
the model based on the quasi-local approach, no assumption on the matter sector is required to recover the Bekenstein-Hawking formula. 
Note however that adapting the notion of fluid approximation to the quasi-local formulation of black holes is straightforward. 

\subsubsection{The fluid approximation}
It is therefore natural to expect that black holes microstates are more likely described as intertwiners between large representations when the black hole is
macroscopic. Hence, we will assume that only these ``large" intertwiners play an important role to understand and to explain some semi-classical properties of black holes. 
As a consequence, we will restrict ourselves to the subspace of such microstates (with large representations) to describe semi-classical
black holes. We dub  these states fluid ``puncels" in analogy with fluid mechanics where a fluid, in the hydrodynamic regime, is understood as a system of interacting fluid parcels.
A fluid parcel is a small amount of fluid which contains a large number of molecules. This is very similar to fluid puncels for black holes  
which are  a small amount of horizon which can be understood as a tensor product of a large amount of more fundamental representations. 
Studying the black hole in terms of fluid puncels instead of more fundamental intertwiners
defines what we call the fluid approximation.

\subsubsection{Puncels}

\bfig
    \centering
    \includegraphics[width=8cm]{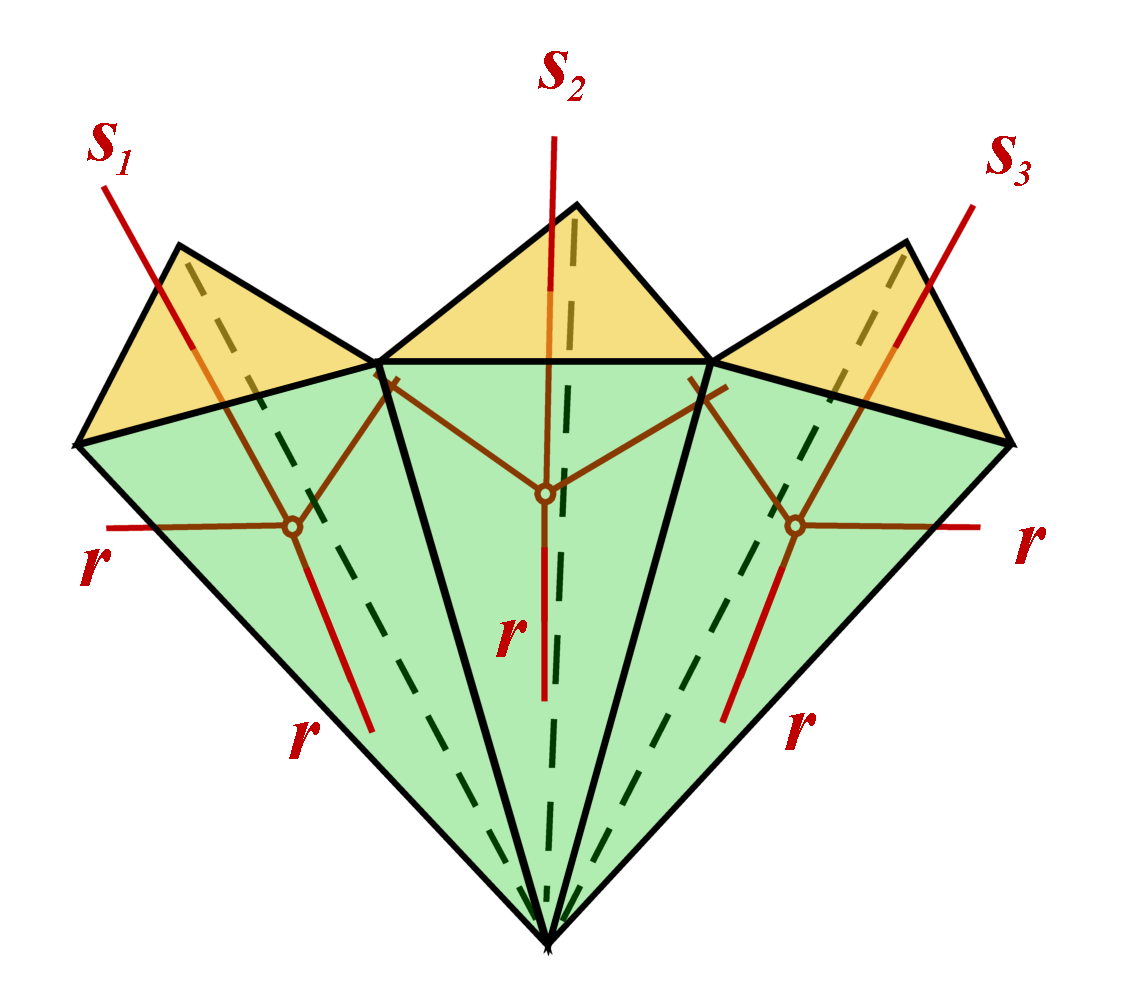}
    \caption{Representation of three puncels of a black hole in fluid approximation. Each tetrahedron is composed by 3 (green) triangles inside the horizon ($\tau^I$) and 1 (yellow)
    traingle on the horizon $\tau^H$.
    Only a part of the triangulation $\Delta$ (black lines) and the corresponding
    dual graph $\Gamma$ (red lines) are represented here.  The dual graph is associated to an intertwiner between the representations $s_1,s_2,s_3,\cdots$.}
    \label{puncels}
\efig
Let us propose a more precise definition of the black hole in the fluid representation. Its microstates 
satisfy the following properties. 
\begin{enumerate}
\item They are intertwiners between  $n$ large representations  $s_i \propto \sqrt{a_H}/\ell_p$ (or $j_i \propto \sqrt{a_H}$ in the quasi-local approach). 
Hence they belong to the space $\text{Int}(\otimes_i V_{s_i} )$.
\item  They are decomposed into $4$-valent intertwiners (inside the horizon) in a way that they are associated to a 4-valent complex $\Gamma$ which is assumed
to be dual to a triangulation $\Delta=\Gamma^*$ of a three dimensional ball whose boundary is the horizon. 
\item Each tetrahedron $T$ of the triangulation $\Delta$ is associated to a 4-valent intertwiner of the form 
\bea \label{symmetrized int}
\iota: V_{s} \otimes V_{r_1} \otimes V_{r_2} \otimes V_{r_3} \rightarrow \mathbb C
\eea
where $s$  colors the face on the horizon and  $r_i$ color all the faces inside the hole. 
\end{enumerate} 
Puncels are  represented in the picture \ref{puncels}. 
As a consequence,  a semi-classical state of a black hole  (in the fluid approximation) with horizon $a_H$ is  determined by a family  of representations 
$(s_\ell,r_{\ell_1}, r_{\ell_2},r_{\ell_3})$ and by 
a family of $n$ intertwiners of the type \eqref{symmetrized int}. The colors $(s_\ell,r_{\ell_1}, r_{\ell_2},r_{\ell_3})$ represent the area of the triangles $\tau_\ell^{H}$ on the horizon and $\tau_{\ell j}^I$ ($j=1,2,3$) inside the hole which are boundaries of the tetrahedron $T_\ell$ according to
\bea\label{triangles}
\text{Area}(\tau_\ell^H) = 4\pi \ell_p^2 s_\ell \, \qquad \text{and} \qquad
\text{Area}(\tau_{\ell_j}^I) = 4\pi \ell_p^2 r_{\ell j} \, .
\eea  
When all the representations $s_\ell$ are equal to $s$ and $r_{\ell j}=r$, we say that the decomposition of the black hole in terms of puncels is fully symmetric.
In that case, the number $n$ of  fully symmetrized puncels of the black hole is trivially given by
\bea
n=\frac{1}{4\pi \sigma} \, \frac{\sqrt{a_H}}{ \ell_p^2} \, .
\eea
When we study the semi-classical properties of the black hole, we will limit ourselves to semi-classical microstates which are close to fully symmetric states.
Only those states which are close to being fully symmetric allow to understand semi-classical properties of black holes.
In that case, we will assume that the colors $r_{\ell i}$ inside the hole are all identical to the same value denoted $r$ with no index in the sequel. 
From now on, we will only consider such states. 

\section{Black Hole spectrum and volume fluctuations}
This part is devoted to introduce and study a model of Hawking radiation using the fluid approximation of black holes. 
In this model, the radiation is closely linked to quantum fluctuations of the euclidean volume operator associated to the black hole.
Indeed,
we proceed as in the approaches based on a calculation of the
path integral: we transform lorentzian metrics into euclidean ones. In the path integral context, this is done using a Wick rotation. In our context, we transform
the geometry inside the horizon by an euclidean geometry. Hence, the euclidean black hole appears as a euclidean three-ball whose boundary is the horizon. 
This (kind of) Wick transformation allows us to interpret the puncels as quantum tetrahedra and the fluid approximation as a discretization of the black hole
in terms of quantum tetrahedra. We will make use of this euclidean transformation to understand some aspects of black holes radiation in  the context of Loop Quantum Gravity.

For this reason, we study, in the first subsection,  the action of the volume operator on black hole microstates in the fluid approximation. Then, in a second subsection, 
we deduce the spectrum of the volume in this limit. Finally, in the last subsection, we relate the volume fluctuations to the black hole radiation. 
In particular, we develop a simple model for the radiation based on the fluid approximation of black holes. This model enables us to show that, if we assume that black holes 
radiate a thermal spectrum, its  temperature $T$ scales necessarily as  the Hawking temperature.

\subsection{Volume of the Euclidean black hole}
At the quantum level, microstates in the fluid approximation are  $n$-valent intertwiners $\iota(s_1,\cdots,s_n)$ between the representations $s_\ell$ 
which is decomposed as the (tensorial)  contraction between $n$ puncels \eqref{symmetrized int}. We can formally write  it as
\bea\label{decomposition}
\iota(s_1,\cdots,s_n) = \langle \! \langle \otimes_{\ell=1}^n \iota_\ell(s_\ell,r,r,r) \rangle \! \rangle_\Gamma
\eea
where the notation $\langle \! \langle  \cdot \rangle \! \rangle_\Gamma$ means that internal  indices are contracted according to the graph $\Gamma$
depicted in figure \ref{puncels}. 

\subsubsection{Volume operator: definition and matrix elements in the fluid approximation}
A priori, only $\iota(s_1,\cdots,s_n)$, such that $j_k=1/2(-1+is_k)$ and $\ell=1/2(-1+i r)$ are integers, are well-defined because they correspond to $SU(2)$ intertwiners. When
 $s_k$ and $\ell$ are real, one could interpret $\iota(s_1,\cdots,s_n)$ as an $SU(1,1)$ intertwiner between representations in the principal series. However,
 such intertwiners need a regularization to be well-defined due to the non-compactness of $SU(1,1)$. Moreover, it is not clear how to compute for instance the action of the volume
 operator $\hat{V}$ (which is intrinsically an $SU(2)$ operator) on  $SU(1,1)$ intertwiners. For these reasons, we define the state \eqref{decomposition} in an indirect
 way: for any operator $\hat{O}$ acting on the space of intertwiners \eqref{space of states} endowed with the scalar product $\langle \cdot \vert \cdot \rangle$, 
 the matrix elements
 \bea
 \langle \iota(s'_1,\cdots,s'_n) \vert \hat{O}  \vert \iota(s_1,\cdots,s_n) \rangle
 \eea 
is defined as the analytic continuation of the corresponding $SU(2)$ matrix element according to the analytic continuation rules \eqref{analytic rules}.

The action of the volume operator on the space of $SU(2)$ $n$-valent intertwiners has been deeply studied in \cite{thiemann}. Its matrix elements which involve (6j)-symbols are very 
complicated coefficients. 
 A part from numerical analysis, such a general formula would be totally useless for our purposes. It is even not obvious that 
 the analytic continuation of these coefficients makes sense. 
 Fortunately, as we are going to argue, we expect the action
 of the volume operator to simplify  in the fluid approximation. 
 
 At first sight,  the space of black holes intertwiners \eqref{decomposition} in the fluid limit is clearly not
 stable under the action of the volume operator $\hat{V}$ and we have
 \bea\label{non stable}
 \hat{V} \vert \iota \rangle = \sum_{\iota'} \langle \iota' \vert \hat{V} \vert \iota \rangle \, \vert \iota' \rangle + \sum_\omega \langle \omega \vert \hat{V} \vert \iota \rangle \, \vert \omega \rangle 
 \eea
 where the first sum runs over an orthonormal basis of \eqref{decomposition} and the second one  runs over intertwiners $\omega$ orthogonal to \eqref{decomposition}. 
 However, we expect that the second term in \eqref{non stable} to be negligible compared to the first one in the fluid approximation. 
The reason is that, as we are going to see, the first term in \eqref{non stable} contains  itself the right classical limit  for the volume operator. Then the rest are only quantum
 corrections to the volume in the fluid approximation. We will neglect them.
 Furthermore, we conjecture that the action of the volume operator simplifies further and 
 can be reduced to the following form
 \bea
 \hat{V} \iota(s_1,\cdots,s_n) = \sum_{j=1}^n \langle \! \langle \iota_1 (s_1) \otimes \cdots \otimes [\hat{V} \iota_j(s_j)] \otimes \cdots \otimes \iota_n(s_n) \rangle \! \rangle_\Gamma
 + \, \varepsilon
 \eea 
 where $\varepsilon$ stands for terms which are also negligible in the fluid approximation. The reason is the same as the previous one. 
 Indeed,  the semi-classical of the first term in the expression of the volume operator  
 \bea\label{sc volume op}
 \hat{V}{}^{(sc)} = \sum_{j=1}^n  \mathbb I \otimes \cdots \otimes \hat{V}_4 \otimes \cdots \otimes \mathbb I
 \eea 
acting on \eqref{decomposition} can be shown to lead to the right classical limit.  We used the notation $\mathbb I$ for the identity and $\hat{V}_4$ for the 4-valent 
volume operator acting on one puncel only.

As a conclusion, we will decompose the volume operator as a direct sum of independent 4-valent volume operators $\hat{V}_4$ acting on each puncel of
the black hole. Obviously, this approximation drastically simplifies the analysis of the volume operator on the black hole microstates
in the fluid limit.

\subsubsection{Volume operator acting on a puncel}
It remains to compute the action of the 4-valent volume operator $\hat{V}_4$ on a puncel $\iota(s_j)$. Instead of starting with the general expression, we first simplify
the expression of the classical volume of a puncel, then we quantize it and only then we compute its action. Notice that, in the following subsection, 
we will start with the general expression of $\hat{V}_4$ and we will show how its action on puncels is related to the calculation of this subsection. 

It is immediate to show that the classical volume $V(s)$ of a flat puncel  is given by
\bea\label{Vs}
V(s) =  {\frac{\sqrt{a_H}}{3\sqrt{4\pi}}} J(s) \sqrt{1-\frac{16 \pi }{3\sqrt{3}} \frac{J(s)}{a_H}}
\eea
where $J(s)=4\pi \ell_p^2 s$ is the area of the face on the horizon. 
We obtained this expression from the Cayley-Menger determinant which gives the volume tetrahedron $V(s)$ from the formula
\bea
[12 V(s)]^2 = \frac{1}{2}\det \left(
\begin{array}{ccccc}
0&1&1&1&1 \\
1&0&x^2&x^2&R^2 \\
1&x^2&0&x^2&R^2\\
1&x^2&x^2&0&R^2\\
1&R^2&R^2&R^2&0
\end{array}
\right)
\eea
where $4\pi R^2=a_H$ and $x$ is the length of the equilateral triangle of area $J(s)$, hence $4J(s)=\sqrt{3} x^2$. 
In the fluid approximation, $J(s)=4\pi\ell_p^2 s \sim \ell_p \sqrt{a_H} \ll a_H$, and then \eqref{Vs} simplifies to
\bea\label{Vs}
V(s) \simeq \frac{\sqrt{a_H}}{3\sqrt{4\pi}} J(s) - \frac{2 \sqrt{4\pi}}{9\sqrt{3 a_H}} J(s)^2 \, .
\eea
The quantization is immediate. Summing over all the puncels  leads to the following expression of the semi-classical volume operator
\bea
\hat{V}{}^{(sc)} = \sum_{j=1}^n \frac{1}{3} \sqrt{\frac{{a_H}}{4{\pi}}} \hat{J}_j - \frac{4}{9} \sqrt{\frac{{\pi}}{{3 a_H}}} \hat{J}_j^2
\eea
where $\hat{J}_j$ is the area operator acting on the horizon face of the puncel associated to $\iota_j(s_j)$. As expected
the states $\iota(s_1,\cdots,s_n)$ diagonalize the volume operator $\hat{V}{}^{(sc)}$ in the fluid approximation, and the corresponding
eigenvalues are
\bea\label{sc eigenvalues}
V^{(sc)}(s_1,\cdots,s_n) = \frac{1}{3 \sqrt{4\pi}} a_H^{3/2} - \frac{64 \pi^2 \sqrt{\pi}}{9 \sqrt{3}} \frac{\ell_p^4}{\sqrt{a_H}} \sum_{j=1}^n s_j^2 \, .
\eea
This eigenvalue gives the volume of the euclidean  flat polyhedron inside the horizon whose faces on the horizon have area $a(s_j)=4\pi \ell_p^2 s_j$.
It is given by the volume of the ball of area $a_H$ given by the first term in the formula \eqref{sc eigenvalues} minus some corrections represented by the second term
in  the l.h.s. of \eqref{sc eigenvalues}. 

Note that, if we do not make any  analytic continuation, we  obtain a formula for $V^{(sc)}$ which  depends on the spins $j_k$ instead
of the continuous representations $s_k$. This formula is easily obtained from the previous one \eqref{sc eigenvalues} by the replacement 
$s_k \rightarrow 2\gamma j_k$:
\bea\label{sc eigenvalues discrete}
V^{(sc)}(j_1,\cdots,j_n) = \frac{1}{3 \sqrt{4\pi}} a_H^{3/2} - \frac{256 \pi^2 \sqrt{\pi}}{9 \sqrt{3}} \frac{\gamma^2 \ell_p^4}{\sqrt{a_H}} \sum_{k=1}^n j_k^2 \, .
\eea
Only the expression of the fluctuations is affected by the analytic continuation.

\subsubsection{Starting from  the full volume operator}
We have just constructed a volume operator in the fluid approximation which is trivially diagonalized by the set of intertwiners \eqref{decomposition}.
However, we know that the full volume operator is a priori a non-diagonal operator acting on intertwiners even in the ``simple" case of 4-valent
nodes. Hence, in this subsection, we make a contact between the eigenvalues \eqref{sc eigenvalues} and the semi-classical limit of the full volume
operator. Because of the decomposition \eqref{sc volume op}, we concentrate only on the 4-valent case. 
We will first show how to define the analytic continuation of the volume operator. Then we will show how it simplifies in the fluid approximation. 

The $SU(2)$ 4-valent volume operator $\hat{V}$ (acting on the space of interwtiners $\iota(j_1,\cdots,j_4)$ between finite dimensional $SU(2)$ representations) 
can be represented as a finite-dimensional matrix. It is simpler to write down formally the matrix $\hat{Q}$ such that $\hat{V}=\vert \hat{Q} \vert^{1/2}$ (see \cite{thiemann} for a more
precise definition of $\hat{Q}$).
 The operator $\hat{Q}$ is hermitian and its non-vanishing matrix elements are
\bea\label{Vol matrix}
a_k \equiv Q_{k-1}^k =  - Q_k^{k-1} = 2i \, (8\pi \gamma \ell_p^2)^3 \, \frac{\Delta(k,A_1,A_2) \Delta(k,A_3,A_4)}{\sqrt{k^2-1/4}}
\eea
with $A_l=j_l+1/2$ and the function $\Delta(a,b,c)$ is the area of a flat triangle of lengths $(a,b,c)$ given by the Heron formula
\bea
\Delta(a,b,c) = \frac{1}{4} \sqrt{(a+b+c)(a+b-c)(a-b+c)(-a+b+c)} \, .
\eea
The dimension of the matrix $\hat{Q}$ and the range of $k$ depend on the spins $j_1, \cdots, j_4$. We want to apply the analytic continuation defined by \eqref{analytic rules} keeping 
$\hat{Q}$ hermitian. This can be done if one maps the integer $k$ to a pure imaginary number $k \rightarrow i\kappa$ where $\kappa \in \mathbb R$. Indeed, with this rule,
the functions $\Delta$ in \eqref{Vol matrix} remain real and the ratio $\gamma^3/{\sqrt{k^2-1/4}}$ is also  real.  Hence $\hat{Q}$ is still an hermitian operator after the 
analytic continuation. At this point, the problem is that it is not clear at all whether
$\kappa$ is an integer or a real number. Unfortunately, the analytic continuation is  too weak an approach to conclude about this aspect. 
It is necessary to understand fully the quantization of gravity
in terms of complex Ashtekar variables to know deeper how the volume operator acts when $\gamma = \pm i$. 

For our purpose, it is not necessary to understand the analytic continuation of the volume operator in detail. 
We first simplify its expression in the fluid limit before performing the analytic continuation. For a regular puncel,
$A_1= A_2 = A_3 = l$ and $A_4=j$ where $l$ and $j$ are large and scale as
\bea
j \sim \frac{\text{Area}(\tau_i^H)}{\ell_p^2} \sim \left( \ell_p^{2}  a_H^{-1} \right)^{-1/2}\, , \qquad
l \sim \frac{\text{Area}(\tau^I_i)}{\ell_p^2} \sim\left( \ell_p^{2}  a_H^{-1} \right)^{-3/4}
\eea
where $\tau^H_i$ and $\tau^I_i$ are  the triangles at the boundary of the puncel $T_i$ which lie respectively 
on the horizon and inside the horizon \eqref{triangles}.
Hence,
\bea
{j}/{l} \sim \left( {\ell_p^2}{a_H^{-1}} \right)^{1/4} \ll 1 \, ,
\eea
and then the label $k$ in \eqref{Vol matrix} belongs to $[l-j ; l+j]$. As a consequence, the matrix elements $a_k$ given in \eqref{Vol matrix} are, up to quantum corrections, all identical
and given by $a_k \simeq a_l$ even if the dimension of $\hat{Q}$ tends to infinity. Finally, one can approximate the matrix $\hat{Q}$ at the semi-classical limit 
around its maximal eigenvalue by the scalar matrix
\bea
\hat{Q}^{sc} \simeq 2 a_l \mathbb I \, ,
\eea
where $\mathbb I$ is the identity acting on the vector space of intertwiners.
One obtains immediately the expression of the volume operator in the fluid approximation. It is also a diagonal and scalar operator whose unique eigenvalue is
\bea
\sqrt{2 a_l} = {\frac{\sqrt{a_H}}{3\sqrt{4\pi}}} (8\pi \gamma j \ell_p^2) \left(1- c \frac{8 \pi }{3\sqrt{3}} \frac{(8\pi \gamma j \ell_p^2)}{a_H}\right)
\eea
where $c$ is a numerical constant.
Now we perform the analytic continuation ($\gamma = \pm i$ and $j=is/2$ in the fluid approximation),  one finds
\bea\label{other Vs}
 {\frac{\sqrt{a_H}}{3\sqrt{4\pi}}} J(s) \left(1- c \frac{8 \pi }{3\sqrt{3}} \frac{J(s)}{a_H}\right)
\eea
As expected, we obtain the same leading order term as in the previous subsection \eqref{Vs}, but the two expressions differ
at the subleading order.  Note however that a precise calculation shows that $c$ is very close to $c=1$. This makes a clear contact between the two different derivations
of the volume operator in the fluid approximation. In the sequel, we will consider  the expression \eqref{Vs} for the volume, i.e. we take $c=1$ in \eqref{other Vs}. 

\subsection{Semi-classical spectrum of the volume}
In a first part, we study the spectrum of the volume operator for a  macroscopic black hole of fixed horizon area $a_H$. This analysis will  be very useful to model
back hole radiation  as proposed in section (\ref{radiation}). 

We can easily notice from \eqref{Vs} and with the constraint \eqref{area} that for a given number of representations $s_i$
the volume is maximal when all the representations are equal  $s_i=s$. This configuration corresponds to the quantum state which approximates the
best the classical black hole. It is the most symmetric state, hence the closest to the spherical symmetry. 
In such case, the volume eigenvalue depends only on $n$ and simplifies to
\begin{equation}
    V^{(sc)}(s)\:=\:  \frac{a_H^{3/2}}{6\sqrt{\pi}}  \left( 1  \:-\: \frac{8 {\pi}}{3\sqrt{3} n } \right)\, \qquad \text{with} \qquad
    s=\frac{a_H}{4\pi \ell_p^2 n}.
    \label{volumeAn}
\end{equation}

We can now study the quantum fluctuations of the black hole in term of its euclidean volume around the most symmetric configuration. 
More precisely, we will slightly change the microscopic configuration from the most symmetric one and we compute the change of the volume
eigenvalue keeping the area fixed. This will provide us with the volume spectrum at the semi-classical limit.

To do so, we will distinguish the cases where the spins are discrete from the cases where the spins are continuous. In the primer situation, it makes sense to consider
a transformation of configuration such that the number of spins is constant but their values change from a fundamental $\pm 1/2$ unit. In the later case, we will consider
a transformation between most symmetric configurations where the number of representations changes. As the spin is continuous and can take any real value, 
such a transformation is allowed. 

\subsubsection{Transformations with constant number of spins}
We assume that the spins are discrete. We consider the transformation with constant number of spins defined by the map
\bea\label{transfo1}
(j,\cdots,j) \mapsto (j+\epsilon_1,\cdots,j+\epsilon_n) \qquad \text{with} \qquad \epsilon_i \in  \mathbb{N} /2 \, . 
\eea
A direct calculation shows that
  \bea
 V^{(sc)}(j+\epsilon_1,...,j+\epsilon_n) \:=\: V^{sc}(j) \:-\: \frac{256 \pi^2 \sqrt{\pi}\gamma^2 \ell_p^4}{9\sqrt{3 a_H}} \left[ \left(2j + 1\right) \sum\limits_{i=1}^n \epsilon_i \:+\:\sum\limits_{i=1}^n \epsilon_i^2 \right]       \, .
  \eea              
For the area to be fixed, we impose the condition $\sum_i \epsilon_i = 0$. 
 Hence, the first gap in the volume is obtained when only two \(\epsilon_i\) is non zero and equal to \(\pm \frac{1}{2}\). By permutation symmetry we can choose \(\epsilon_1\) and \(\epsilon_2\) as the two non zero transformations. The gap in the volume is simply given by
    \begin{equation}
        \Delta V_n^{(1)} \:=\:V^{(sc)}   {(j+\frac{1}{2},j-\frac{1}{2},j,...,j)}   - V^{sc}(j) =  - \frac{64 \pi^2 \sqrt{\pi}\gamma^2}{9\sqrt{3}} \frac{\ell_p^4}{\sqrt{ a_H}} \, .
        \label{deltaV1}
    \end{equation}
    When all the spins are changed by \(\pm \frac{1}{2}\), we obtain a second gap of volume of a different order of magnitude
    \begin{equation}
        \Delta V_n^{(2)} \:=\: - \frac{64 \pi^2 \sqrt{\pi} \gamma^2}{9\sqrt{3}}\frac{\ell_p^4}{\sqrt{a_H}}n \:\propto\:\ell_p^3 \, .
        \label{deltaV2}
    \end{equation}
  As a result, the transformations which keep the number of spins constant change the volume by steps of order \({\ell_p^4}/{\sqrt{a_H}}\) 
    until reaching steps of order \(\ell_p^3\). 

To conclude that the area is fixed in the transformation \eqref{transfo1}, we used the linear spectrum $a_j=8\pi \gamma \ell_p^2 j$. However, as we are considering small
changes in the area, the linear approximation might not be available anymore. Let us see how the area changes when we start with the exact spectrum 
$a_j=8\pi \gamma \ell_p^2 \sqrt{j(j+1)}$ for the faces. A direct   calculation shows that
 \bea
                      \Delta(a_H) & = & a_H(j+\epsilon_1,\cdots,j+\epsilon_n) - a_H (j) \nonumber \\
                      & = &  8\pi\gamma \,\ell_p^2 \left(\frac{1}{j}-\frac{1}{2 j^2}+\frac{1}{2 j^3}-\frac{1}{2 j^4}\right) \sum\limits_{i=1}^n \epsilon_i 
                      \:-\: \frac{\pi\gamma \,\ell_p^2}{j^4} \sum\limits_{i=1}^n \epsilon_i^2 \:+\: O\left(\frac{\ell_p^2}{j^5}\right) \, .
\eea
Obviously, we recover the condition  \(\sum_i \epsilon_i \:=\:0\) 
to have a constant area up to the first order. 
Although, we don't really need to impose \(\sum_i \epsilon_i^2 \:=\:0\). Indeed if we compute the coefficient in front of \(\sum\limits_{i} \epsilon_i^2\) in \(\Delta a_H / a_H\), 
the coefficient is in order of magnitude \({\ell_p^6}/{a_H^3}\), whereas the same coefficient in \(\Delta V / V\) is in order of magnitude 
$\ell_p^4/{a_H^2}$. Then, it is correct to claim that the transformation \eqref{transfo1} keeps the horizon area  unchanged.

 \subsubsection{Transformations with a change of the number of spins} 
 \label{type3}
 Now, we start studying transformations from one most symmetric configuration $(s,\cdots,s)_n$ with $n$ representations $s$ to another symmetric one
 $(s',\cdots,s')_{n+1}$ with one more representation denoted $s'$. The representation $s'$ is changed in such a way that the area $a_H$ remains fixed in this transformation.
 We first suppose that $s$ are continuous such that it is always possible to have such a transformation with $a_H$ fixed. 
From \eqref{sc eigenvalues}, we  directly compute the induced gap of volume and we obtain
    \begin{equation}
        \Delta V_{n\rightarrow n+1} \:=\: \frac{4 \sqrt{\pi}}{9\sqrt{3}}\:\frac{ a_H^{3/2}}{n\left(n+1\right)} \, .
        \label{deltaV3}
    \end{equation}
    As expected, the volume with $n+1$ representation is larger than the volume with one less representation.
    To be more exhaustive we could consider different transformations which change also the number of spins. For instance, $(s,\cdots ,s)_n\rightarrow (s,\cdots s,s',\cdots,s')_{n+1}$ 
    where only \(k\) representations (\(s'\)) are homogeneously changed. 
    In this case we can show that the expression of the gap of volume generalizes
    the previous formula and is given by
    \begin{equation}\label{general vol fluct}
        \Delta V_{k,n\rightarrow n+1} \:=\: \left(1+\frac{1}{k}\right)\frac{4 \sqrt{\pi}}{9\sqrt{3}}\:\frac{ a_H^{3/2}}{n\left(n+1\right)} \, .
            \end{equation}
    We can conclude that the gap is in the same order of magnitude as when \(k\:=\:n\) and that the minimum of gap is obtained when \(k\:=\:n\). We arrive to the same result 
    when the addition of a representation is not homogeneously distributed. 
   Finally, in the fluid approximation and using the relations \eqref{means}, we conclude that
   the transformations which change the number of representations composing the black hole create a difference of volume $\Delta V_n^{(3)}$ with order of magnitude $\,\ell_p^2 \sqrt{a_H}$ as shown in
   \bea\label{gap3}
   \Delta V_n^{(3)} = \frac{4 \sqrt{\pi}}{9 \nu^2 \sqrt{3}}\:\ell_p^2 \sqrt{a_H} \,.
   \eea
If the representation are discrete (spin $j$), there is in general no transformation that sends $n$ spins $j$ to $n+1$ spins $j'$ where both $j$ and $j'$ are integer with the
condition that $a_H$ remains fixed. A part from particular cases, the area of the horizon cannot be fixed and fluctuates. Its fluctuations $\Delta a_H$ can be easily 
bounded from above according to ${\Delta a_H}/{a_H} \leq \ell_p/\sqrt{a_H}$.  Hence, these fluctuations can be larger than the volume fluctuations $\Delta V_n^{(3)}/V^{(sc)} \sim (\ell_p/\sqrt{a_H})^2$.
As a result, the calculation of \eqref{gap3} makes sense only when the representations are continuous, otherwise one cannot assume that the area is unchanged during the transformation.

\subsubsection{Summary}

\bfig[h!]
    \centering
    \includegraphics[width=9cm]{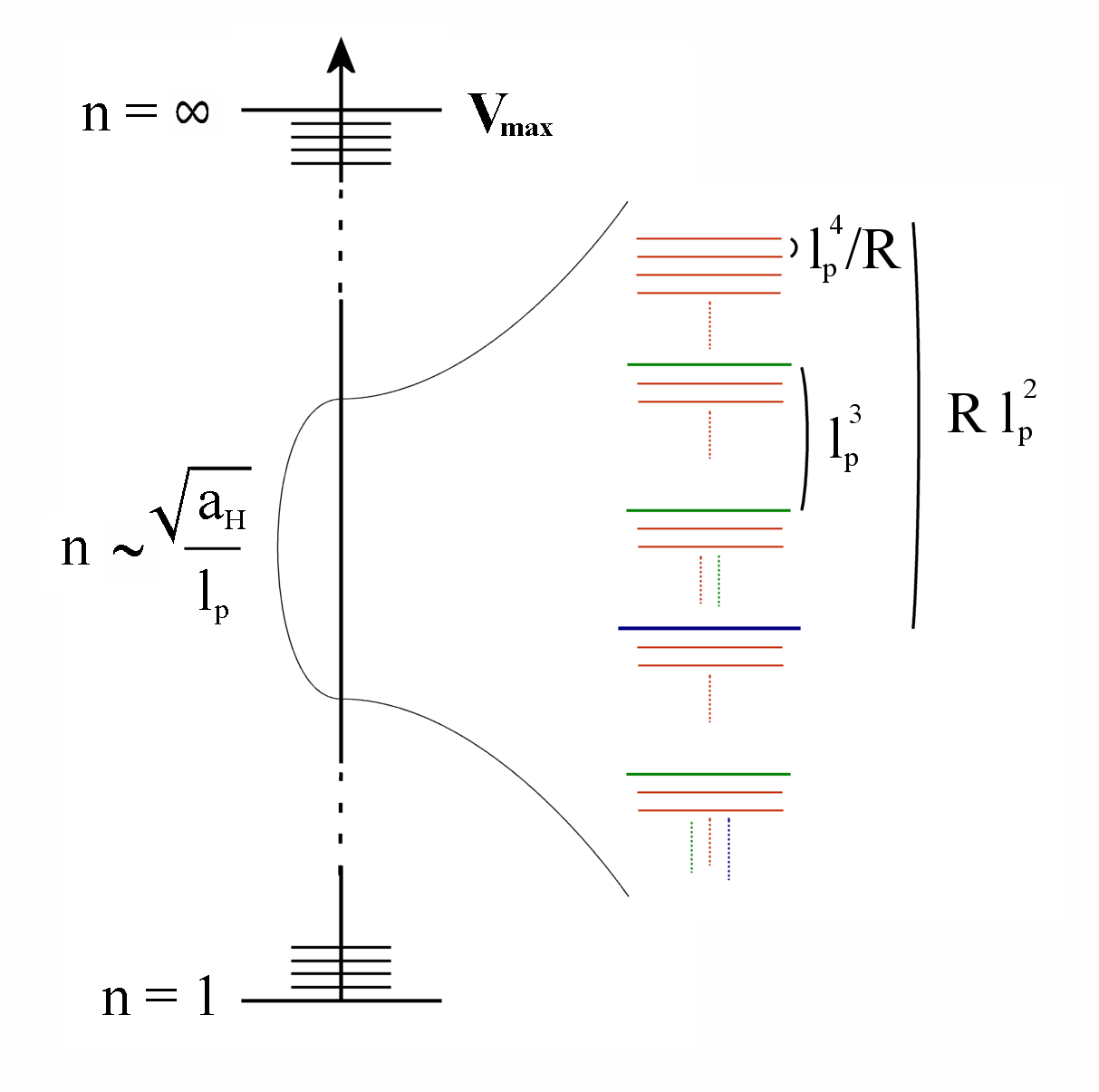}
    \caption{Representation of the spectrum of volume of a black hole in fluid approximation where $n \sim \sqrt{a_H}/\ell_p$. 
    We represented in the same graph fluctuations of the discrete spins model
    and those (which are larger) of the continuous representations model. The notation $R$ is for the Schwarzchild radius. }
    \label{gapofV}
\efig

We gather all the quantum fluctuations of the volume together in the figure  \ref{gapofV}. This graph is instructive because it shows the different scales in the fluctuations even though 
\eqref{deltaV1} and \eqref{deltaV2} concern models where faces are colored with discrete spins whereas \eqref{deltaV3} concerns models with a continuous representation.

We could  also have studied the same type of transformations in the case of an initial configuration  where all the representations are not equal. The computations are heavier but we could arrive 
to the same conclusions: there are three kinds of gap in the volume, the orders of magnitude are the same as the homogeneous case and only the numerical coefficients in the different \(\Delta V\) differ. 

Note that the black hole spectrum in the deep quantum regime (when $n$ is small) has been studied numerically in \cite{spectro1,spectro2} for instance.

\subsection{Black Hole radiation and Hawking temperature}
\label{radiation}
In this last section,  we consider the model of the black hole in the fluid approximation where the puncels are colored with continuous representations. 
As we have just seen, even when the black hole has a fixed area $a_H$, its euclidean volume fluctuates. We suppose that its fluctuations
are due to creation/destruction of puncels which rearrange themselves spontaneously according to the transformations $(s,\cdots ,s)_n\rightarrow (s,\cdots s,s',\cdots,s')_{n+1}$ 
described above  in subsection (\ref{type3}). The volume fluctuations are then  given by  \eqref{general vol fluct} or \eqref{deltaV3} when all representations
rearrange themselves to the same color. 
We are going to show how the expressions of these fluctuations can be used to propose a model of black hole radiation in the framework of Loop Quantum Gravity.

When the black hole is isolated (interacting with nothing), its area is fixed and nothing  could enable us to probe its volume fluctuations.  One could mesure these fluctuations only if the black hole
interacts with a field whose coupling is sensitive to the volume (in the euclidean picture). 
In that case, one can access to a measurement of the fluctuations of the volume measuring changes
in the field states. 
When such a field is coupled to the black hole geometry, it excites the internal structure of the quantum black hole and then the typical energy of the field should be closely 
related to the volume fluctuation of the geometry. 
We are going to show that it is possible to make such a scenario more precise and we will argue that it  could lead us towards  a simple fundamental explanation of the Hawking radiation
 from the point of view of Loop Quantum Gravity. 

\medskip

We start with the expression of the volume fluctuation  \eqref{deltaV3}. From a semi-classical point of view, this volume fluctuation can be understood as resulting from a
quantum deformation of the horizon of the black hole, and more precisely from a local and quantum deformation of its radius (and the area, hence the total energy is kept fixed). 
Due to the quantum fluctuations, the black hole is no more
spherical and the typical length scale of its mean radius fluctuations (with respect to Schwarzchild radius)
is given by
\bea\label{defect fluct}
\overline{\delta R}  \equiv \ell_v = \frac{\Delta V_n^{(3)}}{a_H} \; \sim \; \frac{\ell_p^2}{\sqrt{a_H}} \, .
\eea
This length scale is very tiny in the sense that it is much smaller than the Planck length itself. It is supposed to give the mean deformation of the radius on the whole horizon. 
Because it is small, it can only be explained by a semi-classical configuration of the black hole where the horizon is globally spherical but at some locations there are some
defects. 
These defects have a quantum origine and they slightly modify locally the radius of the black hole. The typical amplitude of these fluctuations is the Planck length $\ell_p$ itself
which is the smallest (hence the more fundamental) possible length. From the description in terms of puncels, this means that most of the puncels are totally identical
but some of them have a defect which locally modify the radius of the horizon with a scale length $\ell_p$. 
More precisely, they should correspond in the fluid approximation
to a configuration $(s,\cdots,s,s',\cdots,s',s'',\cdots,s'',\cdots)$ where different ``small'' paquets of puncels acquire a representation $s',s'',\cdots$ different from $s$.

Let us estimate how many defects are spread on the horizon to reach the mean fluctuation \eqref{defect fluct}. If we denote by $x$ their number, it
satisfies the condition  
\bea
\delta R \, \sim \, \frac{x \ell_p}{n} \, \sim \,  \ell_v \,,
\eea
This formula is easily interpreted as we compute on the l.h.s. the mean of the radius deformation $\delta R$. As $n \sim \sqrt{a_H}/\ell_p$, we immediately
see that $x$ is necessarily of order $x \sim 1$. As a result, there are few defects (number of order 1) spread of the horizon. This has been illustrated in picture (\ref{defects}). Because
of the spherical symmetry, we suppose that the defects form a regular pattern on the horizon.
The typical distance between two defects define a 
wavelength $\lambda$ which characterizes the deformation wave on the horizon. Equivalently, when a field interacts with the black hole, it might be  a scattering of the field
on these defects, hence the field might excite the ``volume" modes 
when its wavelength  scales as $\lambda$.
Here $\lambda \sim \sqrt{a_H}$ because $x \sim 1$ and the 
corresponding temperature is given by
\bea
T \, = \, \frac{h}{\lambda k_B} \, \sim \, \frac{h}{GM k_B}
\eea
where $h$ is the Planck constant, $M$ the mass of the black hole, $G$ the Newton constant and $k_B$ the Boltzmann constant. We recover the scaling
of the Hawking temperature $T_H$
\bea 
T_H = \frac{1}{8 \pi} \frac{h}{ GM k_B} \, .
\eea
The model is not sufficiently precise to obtain the exact expression of the black hole temperature.
\bfig[h!]
    \centering
    \includegraphics[width=5cm]{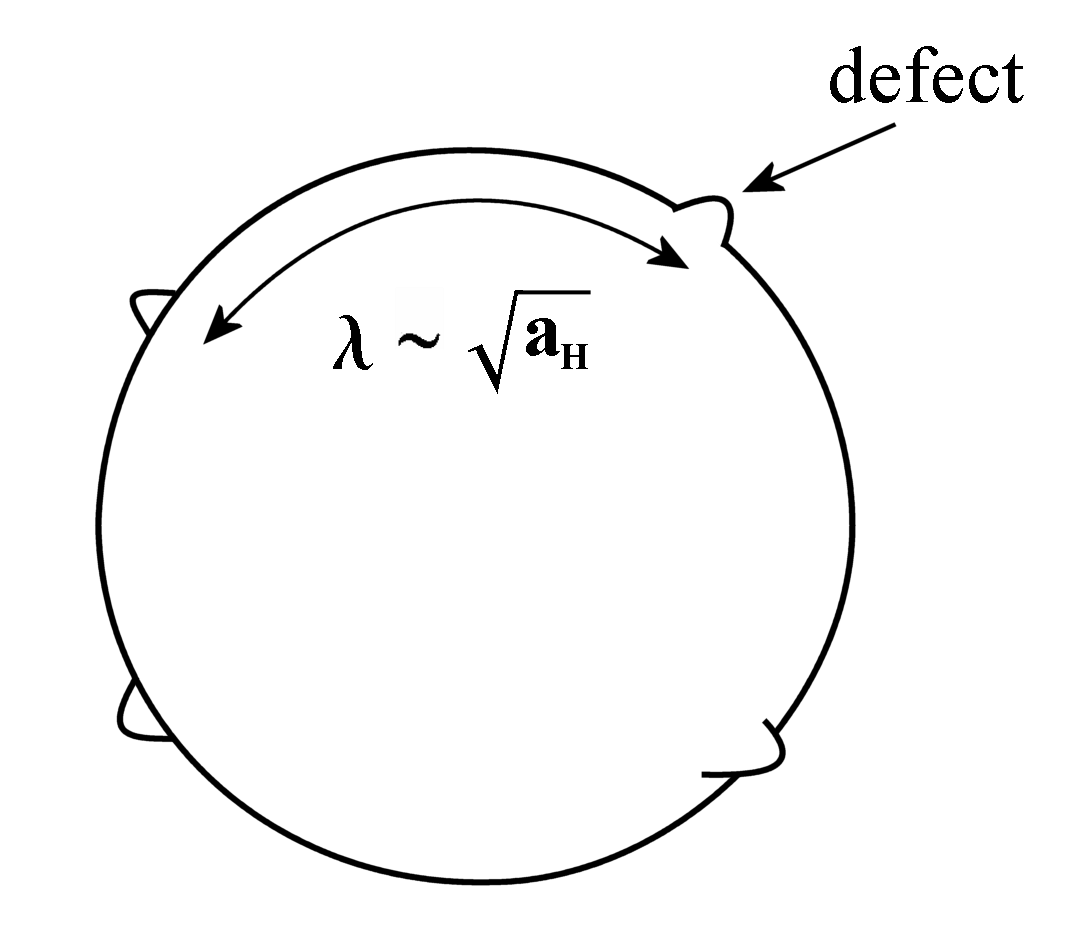}
    \caption{Semi-classical graphical representation of the quantum fluctuation of a black hole. Everything happens as if there were some defects spread on the horizon. These defects form a pattern of typical length $\lambda$. The defects break the spherical symmetry as would do any classical perturbation of the Schwarzchild metric. 
    This length $\lambda$ is supposed to give the typical wavelength of the Hawking radiation of the black hole.}
    \label{defects}
\efig

We conclude this Section with one remark. At the semi-classical limit, we have shown that the volume spectrum is linear and then 
there is a large number of wavelength $\lambda_p$ (or equivalently frequencies $\omega_p$) labelled by an integer $p$ associated to the different volume gaps
and given by $\lambda_p = \lambda/p$ or $\omega_p=p\, \omega$. In the context of black holes, this reminds  us some aspects of quasi-normal modes \cite{QNM} which, at some
classical limit, have been shown to behave in the same way with the difference that they possess also an imaginary part, responsible for the decay of the modes.
The real part of quasi-normal modes frequencies scale exactly as the frequencies we found here. Furthermore, quasi-normal modes are also associated to a deformation 
(from the spherical symmetry) of the black hole horizon. Hence, it would be very interesting to understand whether the analogy between these quantum deformations 
and quasi-normal modes is deeper.

\section{Discussion}
In this article, we propose an idea which could help understanding Hawking radiation from the point of view of Loop Quantum Gravity. It is based on some
physical arguments and interpretations. But it clearly needs to be studied deeper for us to claim that the scenario we are proposing is indeed realistic.
We considered an isolated spherical black hole of fixed area $a_H$. Its microstates are intertwiners and they are obviously all eigenstates of the horizon area 
operator with the same eigenvalue $a_H$. Hence, we need another operator (different from the area) to probe the quantum structure of the black hole. 
We used the euclidean volume operator which indeed raises the degeneracy and allows to probe the internal structure of the microstates.  We studied its spectrum
in the fluid approximation and we interpreted the volume fluctuations in terms of a quantum deformation of the horizon. We argued that
this deformation are described in terms of defects which are regularly spread on the horizon. As the typical length of the defects pattern is $\lambda \sim \sqrt{a_H}$,
there is a natural temperature associated to the quantum deformation of the black hole which scales as the Hawking temperature.  We conclude that Hawking radiation
might originate from quantum fluctuations of the horizon.

However, our model is based on many hypothesis and we made used of the euclidean volume to define quantum fluctuations of the horizon. Many points remain
to be clarified and we hope to study these aspects in the future. In particular, it would be interesting to have a robust description of the interaction of a field
with the black hole in the framework of loop quantum gravity. This could enable us to provide a law energy effective description of the quantum fluctuations and
to write a wave equation where the typical length $\lambda$ would appear explicitely.  
It would also be very instructive to describe the fluid approximation in terms of coherent states,
or twisted geometry. We hope to develop these ideas in the future.

\end{document}